\documentclass[aps,twocolumn,superscriptaddress,floatfix,longbibliography]{revtex4-1}
\usepackage{lipsum}
\usepackage{graphicx}
\usepackage[export]{adjustbox}
\usepackage{amsmath,amssymb,wasysym}
\usepackage{braket}
\usepackage{mathtools}
\usepackage{mathrsfs}
\usepackage{verbatim}
\usepackage{makecell}
\usepackage{cases}
\usepackage{bbm}
\usepackage{enumitem}
\usepackage[dvipsnames]{xcolor}
\usepackage{hyperref}

\newcommand{\im}{\mathbf{i}}

\newcommand{\abs}[1]{\left| #1\right|}

\newcommand{\BigO}{\mathcal{O}}

\def\average#1{\left< #1 \right>}

\begin{document}
	\title{Level Set Percolation in Two-Dimensional Gaussian Free Field}
	\author{Xiangyu Cao}
	 \affiliation{Laboratoire de Physique de l'Ecole Normale Sup\'erieure, ENS, Universit\'e PSL,
CNRS, Sorbonne Universit\'e, Universit\'e de Paris, 75005 Paris, France}
\author{Raoul Santachiara}
 \affiliation{Universit\'e Paris-Saclay, CNRS, LPTMS, 91405, Orsay, France}
	\begin{abstract}
       The nature of level set percolation in the two-dimension Gaussian Free Field has been an elusive question. Using a loop-model mapping, we show that there is a nontrivial percolation transition, and characterize the critical point. In particular, the correlation length diverges exponentially, and the critical clusters are ``logarithmic fractals'', whose area scales with the linear size as $A \sim L^2 / \sqrt{\ln L}$. The two-point connectivity also decays as the log of the distance. We corroborate our theory by numerical simulations. Possible conformal field theory interpretations are discussed. 
	\end{abstract}
	\date{\today}
	\maketitle
		
\noindent\textit{Introduction.}-- Imagine a random landscape being flooded with water. As the water level rises, initially disconnected lakes connect with one another and form eventually an infinite ocean. Does that happen when the flooded area reaches a critical density? If yes, what are the critical properties of the transition? These are the basic questions of the percolation theory of random fields --- the relief profile is given by a field $\phi(x)$ and the flooded area $\{x: \phi(x) \le h \}$ is known as the level set (or excursion set) of height $h$~\cite{efros}. Such questions arise naturally in topography and planet science~\cite{percolation_rev1,kalda_coastline}, but also in transport properties of disordered systems~\cite{zallen71,trugman83,shklovskii2013electronic}, and have been extensively studied (for recent reviews, see \cite{percolation_rev2,saberi15review}). 

\begin{figure}
    \centering
    \includegraphics[width=.98\columnwidth]{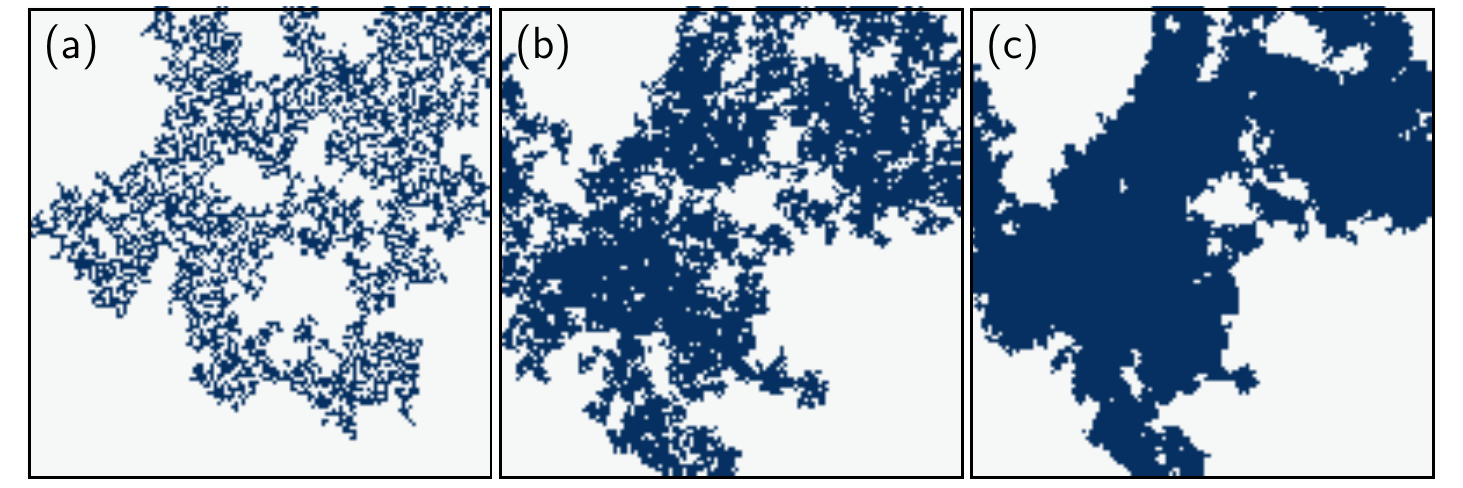} 
    \caption{Level set percolation in random fields with different roughness exponents. \textbf{(a)} When $H = -1 < 0$, there is a percolation transition, with fractal clusters appearing at  criticality. \textbf{(b)} When $H = 0$ (2D GFF), the critical clusters are logarithmic fractals. \textbf{(c)} When $H = \frac12 > 0$, there is no sharp transition, and the clusters are not fractal.}
    \label{fig:three}
\end{figure}

The answer to these questions depends crucially on the statistical properties of the field. If it is short-range correlated, there is a second-order percolation transition~\cite{Molchanov1983,beffara,rivera2019critical}: in the thermodynamic limit, infinite clusters (connected components) of the level set never appear below some critical threshold, and always appear above it. The critical point is in the universality class of standard uncorrelated percolation. For long-range correlated fields characterized by a single Hurst roughness exponent $H$, 
$ \average{\phi(x) \phi(y)}\sim |x-y|^{2H}$, the situation is richer.
When $H < 0$, a transition still exists. At criticality, the infinite clusters are fractals, see Fig.~\ref{fig:three}-(a). Their geometric properties depend on $H$, and have been analytically and numerically characterized~\cite{weinrib84longrange,shrenk15,Janke_2017,nina}. When $H > 0$, there is no sharp transition~\cite{Schmittbuhl_1993}, and the clusters are ``compact'' objects instead of fractals, see Fig.~\ref{fig:three}-(c). 

\begin{figure}
    \centering
    \includegraphics[width=.95\columnwidth]{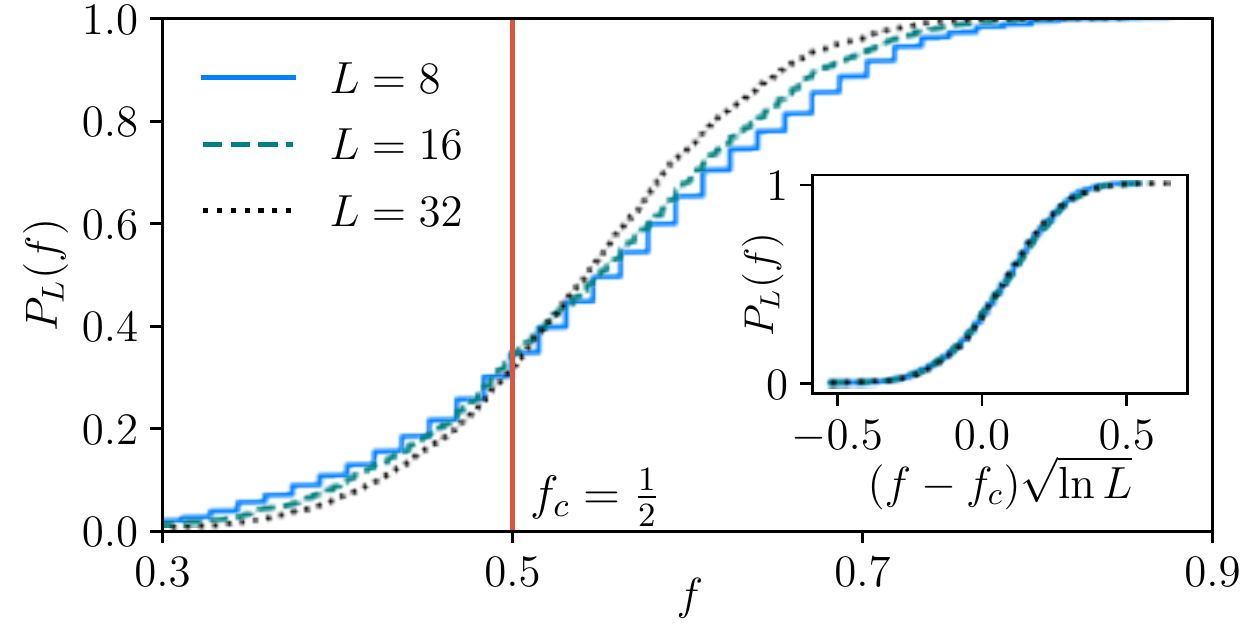}
    \caption{Percolation threshold of 2D GFF level sets on square lattices of size $L \times L$. \textbf{Main}: The existence probability of a percolating cluster $P_{L}(f)$, where $f = (\text{level set area}) / L^2$ is the level set's mean density. \textbf{Inset}: Finite-size scaling collapse according to the correlation length prediction \eqref{eq:corr_length}, identifying $\xi = L$. The critical value $f_c = 1/2$ regardless of lattice details. 
    All numerics in this paper are done with {real-valued} 2D GFFs on periodic square lattices, generated by the standard Fourier filter method~\cite{fourier,supp}.
    }
    \label{fig:pc}
\end{figure}

A natural question is then what happens at $H=0$, where $\phi$ is log-correlated. This is arguably the most interesting point, especially in two dimensions, as it corresponds to the 2D Gaussian Free Field (GFF). A simple model of elastic interfaces~\cite{Aarts847}, its importance in low-dimensional physics~\cite{giamarchi2003quantum}, 2D critical phenomena~\cite{dofa_npb84} and random geometry~\cite{duplantier09prl} cannot be overstated. Despite the enormous amount of studies on the 2D GFF, the percolation of its level sets has been little discussed (by contrast, a number of rigorous results exist for the GFF in $D \ge 3$ dimensions~\cite{bricmont,drewitz} {  or for the GFF on a transient tree ~\cite{Ab_cherli_2018}}). There are indeed a few arguments for dismissing this problem as uninteresting. For example, since the correlation length exponent $\nu$ diverges as $H$ approaches zero from below~\cite{weinrib84longrange}, there would be no transition at $H = 0$~\cite{percolation_rev1} (see however \cite{Schmittbuhl_1993}). Moreover, even if there is a transition, it could be trivial from a geometric point of view~\cite{LEBOWITZ1986194}, since the fractal dimension of the critical clusters approaches $D_{\text{f}}=2$ in the same limit~\cite{Janke_2017,nina,sepulveda2019}.
However, these conclusions can be challenged by looking at some large level set clusters of the 2D GFF, shown in Fig.~\ref{fig:three} (b). They are apparently distinct from their $H > 0$ as well as $H < 0$ counterparts. Numerical simulations also indicate a sharp transition at a critical density $1/2$ in the thermodynamic limit, see Fig.~\ref{fig:pc}. Could the above arguments have missed something subtle?

In this Letter, we revisit the problem of level set percolation in the 2D GFF. By an analytical argument based on the loop-model reconstruction of the 2D GFF, we show that its level set percolation is a nontrivial critical phenomenon, characterized by an exponentially diverging correlation length. At criticality, we find that the area of the large clusters with linear size $L$ scales as 
\begin{equation}
     A \sim \frac{L^{2}}{\sqrt{\ln L }} \,. \label{eq:fractal}
\end{equation} 
So, the clusters have the same fractal dimension $2$ as compact Euclidean forms, but differ from them by a log correction. Such geometric objects have been named ``log fractals''~\cite{mandelbrot,falconer2004fractal,INDEKEU1998294}. A remarkable example of log fractals is the set of points visited by a random walk on a 2D lattice, which satisfies $A \sim L^2 / \ln L$~\cite{dvoretzky1951,kundu}. We will also show that the probability that two points $x,y$ belong to the same level cluster has a peculiar log decay:
\begin{equation}
    P_{2}(x,y) \approx C - \frac1\pi \frac{ \ln \abs{x-y}}{\ln L} \,, \label{eq:p2}
\end{equation}
where $C$ is an unknown constant and $L$ is the system size. 

\noindent\textit{GFF by loop model.}-- The 2D GFF can be defined by the action 
\begin{equation}
    \mathcal{S}[\phi] = \frac{g}{4\pi} \int \left[\nabla\phi(x)\right]^2 \mathrm{d}^2 x \label{eq:action}
\end{equation}
(the probability of $\phi$ is proportional to $e^{-\mathcal{S}[\phi]}$) where $g$ is the coupling constant, which also determines the log correlation of $\phi$, 
 \begin{equation}
     \average{\phi(x) \phi(y)} = - \frac1{g} \ln  \abs{x-y} \,. \label{eq:twopoint}
 \end{equation}
 Its short-distance divergences need to be regularized in some way. Here, we shall do this by using the $\mathrm{O}(n=2)$ loop model on the honeycomb lattice~\cite{cardy-ziff}, as reviewed below. We expect the critical properties of the level set percolation do not depend on the choice of regularization. 
 
\begin{figure}
    \centering
    \includegraphics[width=\columnwidth]{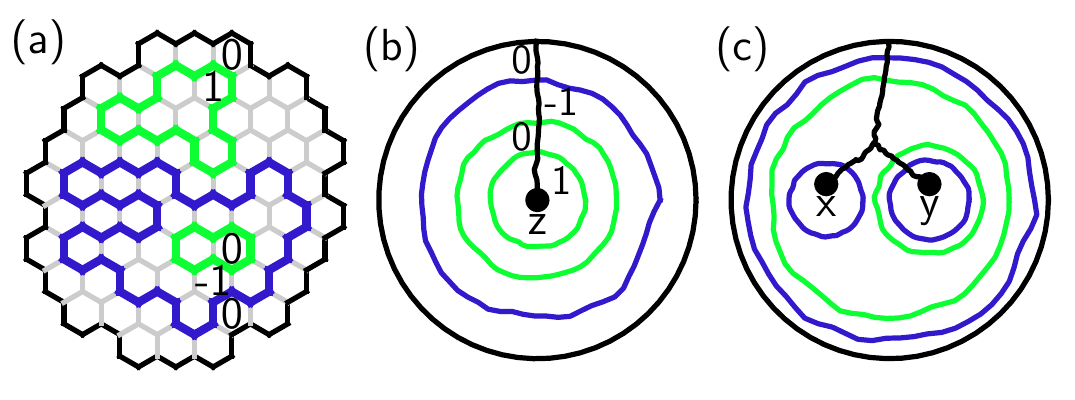}
    \caption{\textbf{(a)} A loop model configuration on a honeycomb lattice and the associated height field $\phi$ (heights are in unit of $\pi$). The color of a loop (blue/green) determines the increment of $ \phi$ across it; $\phi$ is constant in regions between loops. \textbf{(b)} A point $z$ encircled by $d_z = 4$ loops, and a path from the boundary to $z$ (without crossing other loops that are not drawn). The field along the paths gives a random walk realization $\{Z_j\} = 0, -1, 0, 1$. \textbf{(c)} Two points $x$ and $y$ satisfying $q = 2$, $d_x = 3$, $d_y = 4$, connected to the boundary by a branching path. The random walk realization  $\{X_j \} = 0,-1,0,-1$ and $\{Y_j \} = 0, -1, 0, 1,0$ share the first $q$ steps. }
    \label{fig:loops}
\end{figure}

The $\mathrm{O}(n)$ loop models can be defined by a partition function which sums over all configurations $\mathcal{C}$ of disjoint loops on the lattice, see Fig.~\ref{fig:loops}-(a):
\begin{equation}
    \mathcal{Z} = \sum_{\mathcal{C}} n^{N_\mathcal{C}} K^{L_\mathcal{C}} \,,
\end{equation}
where $n$ and $K$ are fugacities associated with the number of loops $N_\mathcal{C}$ and the their length $L_\mathcal{C}$, respectively. When $n = 2$, for any $K \ge K_c = 1/\sqrt{2}$, the model is critical and described by a conformal field theory (CFT) with central charge $c = 1$, see e.g. \cite{estienne2015correlation,Jacobsen2012}. Given a loop configuration, we can assign a random orientation to each loop, and define a height function $\phi$ on the lattice faces such that the loops are its oriented contour lines, with a step $\pm \pi$ across each loop. Assuming the Dirichlet boundary condition for $\phi$ in a simply connected domain, {  that is, using $\phi\vert_{\text{boundary}} = 0$ as the starting point,} the height configurations are in one-to-one correspondence with oriented loop configurations~\footnote{Being simply connected is important here. On a torus, non-contractible loops could lead to a defect, so the field $\phi$ must be compactified~\cite{estienne2015correlation}}.  
It is well known that the scaling limit of the height function is a 2D GFF with a coupling constant satisfying 
\begin{equation}
    n = -2 \cos(\pi g) \stackrel{n=2}\Rightarrow g = 1 \,.
\end{equation}
The above mapping is the first step to the Coulomb gas approach, that has been applied for studying certain random sets of the GFF, notably the loops and the regions between them~\cite{Saleur_89_interloop} (see  \cite{werner2020lecture} and references therein for rigorous works). However, the Coulomb gas approach fails to capture the level cluster, but we can still use the mapping to study them.

\noindent{\it Application to percolation.}-- 
Now that we have a 2D GFF $\phi$ regularized on simply-connected domain with Dirichlet boundary condition, let us consider the level set $\{\phi \le h\}$ for $h > 0$. The \textit{infinite cluster} is defined as the set of points that are connected to boundary by a path in the level set. Note that the \textit{gasket}, i.e., the region not encircled by any loop, is a subset of the infinite cluster. Now, let $z$ be inside a hexagon, the following probability
\begin{equation}
    P_{\infty}(z,h) := \mathrm{Prob}(z \in \text{the infinite cluster})  \label{eq:pzh-def}
\end{equation} 
is the order parameter in percolation theory. It encodes in particular the basic critical exponents. 

A little thought shows that this has much to do with the number $d_z$ of loops that encircle $z$. Indeed, consider a path from the boundary to $z$ that only crosses the $d_z$ loops encircling $z$, see Fig.~\ref{fig:loops}-(b). $z$ is in the infinite cluster if and only if $\phi \le h$ along this path.
By the loop-model construction, the values of $\phi$ along the path, denoted $\{Z_j\}_{j=0}^{d_z}$, form an unbiased 1D random walk with steps $\pm \pi$
\begin{equation}
    Z_0 = 0 \,,\, Z_j - Z_{j-1} = \pm \pi\,,\, j= 1, \dots, d_z  \,.\label{eq:randomwalkZ}
\end{equation}
A classical result states that this random walk never goes above $h$ with the following probability (see e.g.~\cite{kay}):
\begin{equation}
\mathrm{Prob}(\forall j \le d_z ,  Z_j \le h ) \approx \mathrm{erf}\left( \frac{h}{\pi \sqrt{2 d_z}} \right) \theta(h) \,, \label{eq:phzd}
\end{equation}
provided $d_z \gg 1$.  Since $\mathrm{Prob}(\forall j \le d_z ,  Z_j \le h )$ is the probability \eqref{eq:pzh-def} {conditioned} on the random number $d_z$, {  the order parameter can be obtained by averaging over $d_z$. 
In the scaling limit, this can be done by simply treating $d_z$ as deterministic and replacing it by its mean value:
\begin{equation}
    P_{\infty}(z,h) \approx \mathrm{Prob}(\forall j \le \left<d_z\right>,  Z_j \le h ) \,.
\end{equation}
To justify this, observe first that the mean value of $d_z$ is related to the variance of $\phi(z)$:
\begin{equation}
    \average{d_z} = \frac1{\pi^2}  \, \average{\phi(z)^2} \approx \frac1{\pi^2} \ln L \,, \label{eq:mean-d}
\end{equation}
where $L$ is the lattice size, for any $z$ far from the boundary (in lattice units). Meanwhile, the fluctuations of $d_z$ are of order $\sqrt{ \ln L}$~\cite{kesten97,schramm09}, so can indeed be neglected compared to $ \average{d_z} $~\cite{supp} as $L\to \infty$. Combining \eqref{eq:phzd} to \eqref{eq:mean-d}, we have 
\begin{align}
 P_{\infty}(z,h) \approx \mathrm{erf}\left(\frac{h}{\sqrt{2 \ln L}}\right)  \approx \frac{h}{\sqrt{\frac\pi2 \ln L}}\,, \label{eq:pzh}
\end{align}
where we linearized $\mathrm{erf}(x)$ around $x = 0$. } Therefore \eqref{eq:pzh} is valid in the scaling limit $L \to \infty$ with fixed $h > 0$ and $z$ far from the boundary.

With the result~\eqref{eq:pzh} at hand, we can derive most of the claims in the Introduction. Eq.~\eqref{eq:fractal} is immediately obtained by summing \eqref{eq:pzh} over the $\sim L^2$ lattice points. The nature of the log fractal is manifest in the fact that the probability of a point belonging to it decays logarithmically in $L$, as opposed to algebraically in a usual fractal. The order parameter exponent $\beta$ can be also obtained. Introducing the mean density of the level set,
\begin{align}
     f &= \mathrm{Prob}(\phi(z) \le h) \nonumber   = \int_{-\infty}^h e^{-{x^2}/(2{\ln L})} \frac{\mathrm{d} x}{\sqrt{2\pi \ln L}}   \\   & \approx \frac12 + \frac{h}{\sqrt{2\pi \ln L}}  \label{eq:fandh}
\end{align}
for $h \sim \BigO(1)$ and $L \to \infty$, $\beta$ is defined by the non-analyticity of $P_{\infty}$:
\begin{equation}
     P_{\infty}(f > f_c) \sim  (f-f_c)^{\beta}   \,.\label{eq:def_beta}
\end{equation}
Comparing \eqref{eq:pzh}, \eqref{eq:fandh}, and \eqref{eq:def_beta}, we have
\begin{equation}
\beta = 1  \,,\label{eq:beta}
\end{equation}
as well as $f_c = 1/2$; the latter is observed numerically, see Fig.~\ref{fig:pc}. Finally, comparing $P_\infty\sim (f-f_c)^1$ and $P_{\infty} \sim 1/\sqrt{\ln L}$~\eqref{eq:pzh}, and assuming that close to the critical point, the correlation length $\xi \sim L$, we find that $\xi$ diverges exponentially near criticality such that 
\begin{equation}
   |f - f_c| \sim \frac1{\sqrt{\ln \xi}} \,. \label{eq:corr_length}
\end{equation} 
This is in nice agreement with numerical simulations shown in Fig.~\ref{fig:pc}, and explains why the transition sharpens extremely slowly as the system size increases. We note that in the thermodynamic limit, any value $h = \BigO(1)$ is critical, even if $h \le 0$ (the restriction $h > 0$ above has to do with the specific setup of the Dirichlet boundary condition and the definition of the infinite cluster).

It is useful to contextualize the above results as the limit of level set percolation with $H < 0$. It was shown by an extended Harris criterion~\cite{weinrib84longrange} that when $H \in (-3/4, 0)$, the correlation length exponent $\nu(H) = -1/H$. The level cluster's fractal dimension $D_\text{f}(H)$ is not known exactly, yet numerics~\cite{Janke_2017,nina} suggest that $D_\text{f}(H \to 0) = 2$. Our analysis explains these limiting behaviors: $\nu({H=0}) = \infty$ because the correlation length diverges faster than any power law, and $D_{\text{f}}(0) = 2$ because we have a log fractal. Moreover, assuming that $\beta$ is continuous at $H = 0_-$,  we have by hyperscaling
\begin{equation}
    D_{\text{f}}(H) = 2 - \frac{\beta(H)}{\nu(H)} =  2 + H + \BigO(H^2) \,,\, H < 0 \,.\label{eq:DfH}
\end{equation}

We also remark that the boundaries of the level cluster, i.e., the loops, are described by the Schramm-Loewner evolution, SLE$_{\kappa = 4}$~\cite{schramm09GFF,schramm13}, and have a fractal dimension $3/2$~\footnote{Similarly, the boundary of the log-fractal set visited by a 2D random walk is described by another SLE$_{\kappa=8/3}$~\cite{du06}.}. This fractal dimension corresponds to a primary operator  in the CFT of the $\mathrm{O}(2)$ loop model. The same CFT also describes the interior regions between loops, which have a fractal dimension $15/8$. However, the level cluster does not seem to be described by any known CFT. It is a very different object, depending on the ``random topology'', rather than the random geometry, of the loops.

\noindent{\textit{Two-point connectivity.}--} To further illustrate this point, let us consider another standard observable in percolation: the two-point connectivity $P_2(x,y)$ at criticality, the probability that $x$ and $y$ are in the same connected component of the {  level set $\{\phi \le h\}$, for any fixed $h = \BigO(1)$}. This observable has been shown recently to probe very subtle universal properties of critical clusters \cite{nina,jps19two}. Similarly to $P_{\infty}$ above,  $P_2(x,y)$ can be calculated by conditioning on the loop configurations. More precisely, we condition on $d_x, d_y$, and also $q$, the number of loops that encircle {both} points, see Fig.~\ref{fig:loops}-(c). The random walks $X_0, \dots, X_{d_x}$, and $Y_0, \dots, Y_{d_y}$ that record the field value evolution from the boundary to $x$ and $y$ have the same first $q$ steps:
\begin{equation}
X_j = Y_j \,,\, j = 0, \dots, q \,,
\end{equation}
while the remaining $d_x - q$ and $d_y - q$ steps are independent (we have a branching random walk). Then, it is not hard to see that the two-point connectivity conditioned on $d_x, d_y$ and $q$ is the probability that the ``forked'' part of the branching random walk never goes above {  $h$:
\begin{equation}
     P_{2}(d_x , d_y, q) = \mathrm{Prob}(\forall i, j \ge q, X_i \le h, Y_j \le h) \,.
\end{equation}}
The common part ($< q$) is not further constrained since we do not require $x,y$ to belong to the infinite cluster. Similarly to \eqref{eq:phzd} above, we have { 
\begin{align}
  &   P_{2}(d_x , d_y, q) \nonumber \\
    \approx  & \int_{0}^{\infty}    \mathrm{erf}\left(\frac{u \sqrt{q} + h}{\sqrt{2 s_x}}\right)
     \mathrm{erf}\left(\frac{u \sqrt{q} + h}{\sqrt{2 s_y}}\right)  
       \frac{ e^{-\frac{u^2}{2}} \mathrm{d} u}{\sqrt{2\pi}}
\end{align}
as long as $s_x := d_x - q, s_y := d_y - q$ and $q$ are all much larger than $1$. In that limit $h = \BigO(1)$ can be also neglected.} Like $d_x$ and $d_y$, the mean value of $q$ is fixed by the covariance of the GFF: 
\begin{equation}
\average{q} = \frac1{\pi^2} \average{\phi(x) \phi(y)} \approx \frac1{\pi^2} \ln \frac{L}{|x-y|} \,,
\end{equation}
and $q$ also becomes deterministic when $L /|x-y| \gg 1$~\cite{supp}. As a result, we find  
\begin{align}
& P_{2}(x,y) \approx  \int_{0}^{\infty}   \mathrm{erf}\left(\frac{u \sqrt{\ln \frac{L}{|x-y|}}}{\sqrt{2 \ln |x-y|}}\right)^2
 \frac{ e^{-\frac{u^2}{2}} \mathrm{d} u}{\sqrt{2\pi}} \,, \label{eq:P2_Dirichlet}
\end{align}
valid in the scaling limit: $1 \ll |x-y| \ll L$, with $x,y$ far from the boundary. In the regime where $|x - y| \gg \sqrt{L}$,  the error functions can be linearized, and \eqref{eq:P2_Dirichlet} simplifies to \eqref{eq:p2} in the Introduction, with the offset predicted as $C = 1/\pi$.

\begin{figure}
    \centering
    \includegraphics[width=.95\columnwidth]{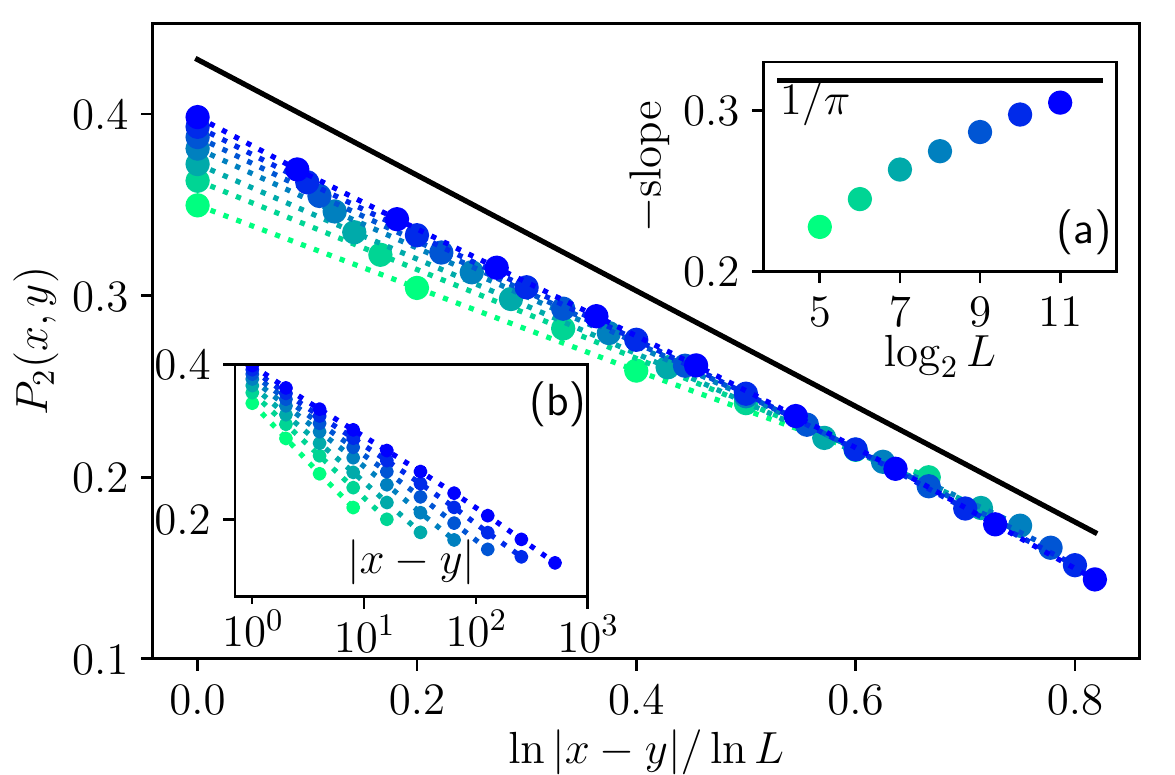}
    \caption{Testing the two-point connectivity prediction~\eqref{eq:p2} with $h =0 $ level sets of the 2D GFF on periodic square lattices of size $L\times L$. \textbf{Main}. The averaged two-point connectivity $P_{2}(x,y)$ as a function of $\ln|x-y|/\ln L$. The dots are numerical measure with $L = 32, 64, \dots, 2048$; see inset (a) for color code. The solid line represents the prediction \eqref{eq:p2} (with the exact slope $1/\pi$ and arbitrarily adjusted offset $C$). \textbf{Inset (a)} {  The slope as a function of system size, obtained by a global linear fit of each curve.} A convergence to the prediction $1/\pi$ as $L\to \infty$ is observed. \textbf{Inset (b)} The raw data. Plotting $P_2$ against $|x-y|$ as usual, we observe no collapse. The dotted lines are a guide to the eye. }
    \label{fig:p2}
\end{figure}

We now test \eqref{eq:p2} numerically, see Fig.~\ref{fig:p2}. The results confirm nicely the $ \ln |x-y| / (\pi \ln L)$ dependence on the distance predicted in \eqref{eq:p2}, including the exact prefactor and throughout the scaling regime. On the other hand, the offset $C$ does not agree with the analytical prediction. The reason of this is two-fold. First, we recall that Eq.~\eqref{eq:P2_Dirichlet} is derived on a simply connected domain with Dirichlet boundary condition, while the numerics is performed on a torus (see caption of Fig.~\ref{fig:pc}). Moreover, we are \textit{a priori} far from the thermodynamic limit. Indeed, our analytical argument relied on a large average number of encircling loops. Yet, according to \eqref{eq:mean-d}, there are in average $0.8$ loops encircling a point in a lattice of $L = 2048$ (even on an Avogadro-scale lattice with $L = 10^{10}$, there would be $2.3$ loops in average). The offset is affected by the different infrared regularization, and the abundance of realizations with few loops. In contrast, the $\ln |x-y| / (\pi \ln L)$ dependence appears to be remarkably robust. Therefore, we propose \eqref{eq:p2}, with an unknown offset $C$, as a general prediction of two-point connectivity.

The log in \eqref{eq:p2} can be explained by a rather simple argument. Indeed, for $H<0$, the level clusters are standard fractals, with the two-point connectivity decaying as a power law $P_2(H) \sim |x-y|^{2 (  D_{\text{f}}(H)-2)}$~\cite{Kapitulnik_84}. By \eqref{eq:DfH}, $P_2(H) \sim 1 + 2H\ln|x-y|$ for $H$ close to $0$ and $\ln |x-y| \ll 1/(-H)$. {  Now, in this regime, a field with $H < 0$ is indistinguishable from a 2D GFF in a system of size $L$ such that $\ln L = 1/(-H)$~\cite{supp}. Therefore we can rewrite $ P_2(H) \sim 1 - \ln |x-y| / \ln L$, recovering the form of \eqref{eq:p2}.}  

\noindent\textit{Conclusion.}-- We showed that the level sets of the 2D Gaussian Free Field have a nontrivial percolation transition, and outlined a theory of the critical point. In particular, the critical level clusters are found to be log fractals, whose connectivity properties are determined by the random topology of the contour lines. The analysis presented above can be extended to any $n$-point connectivity, which is mapped to a branching random walk. The emergence of such a hierarchical structure in the 2D GFF is not at all new. For instance, it is crucial in the problem of extreme value statistics of the 2D GFF, also known as log-correlated random energy models, or multiplicative chaos~\cite{wen,carpentier,Fyodorov_2008,rhodes2014,madaule2016}. The latter problem admits, nevertheless, a conformal field theory description~\cite{lft,lft2,remy2020}. Whether the same can be said of the level clusters of the 2D GFF seems to be an interesting question. We remark that the level clusters for $H  < 0$ seem to be described by a new CFT, some features of which have been numerically studied~\cite{nina}. As discussed above, the logs that appear in \eqref{eq:p2} can be obtained from certain correlators in the $H < 0$ CFT, which do not involve an indecomposable representation of the conformal symmetry~\cite{savi13,nivesvivat2020logarithmic}, or a continuous spectrum~\cite{ribault2014conformal}. Whether such structures exist in the putative $H = 0$ CFT remains to be seen.

We thank Sebastian Grijalva, Nina Javerzat and  Alberto Rosso for collaborations on related questions, and Malte Henkel, Pierre Le Doussal, Satya Majumdar, Sylvain Ribault and Hugo Vanneuville for helpful discussions.

	\bibliography{refs}
	
	\begin{widetext}
		\subsection{Numerical Methods}
	
	\textbf{Gaussian Free Fields.} 
	The numerical tests are all performed on a square lattice of size $L \times L$, with periodic boundary conditions on both directions, $x \equiv x +L$, $y \equiv y + L$.  The 2D GFF can be numerically generated by the standard Fourier filter method:
	\begin{equation}
	    \phi(x, y) := \mathrm{Re} \left( {\sum_{\substack{p,q=1,\dots,L \\ (p,q) \ne (0,0)}}}
	    e^{\frac{2\pi}L \im (q x + p y )} 
	    \left( \sin\left(\frac{q \pi }L\right)^2 + \sin\left(\frac{p \pi }L\right)^2  \right)^{-\frac{1 + H}2} \mathcal{N}_{p,q} \right) \,,\, x, y = 1, \dots, L \,.
	\end{equation}
	Here, $H = 0$ gives us the 2D GFF, while $H \ne 0$ corresponds to fractional GFF with a Hurst exponent $H$. $\mathcal{N}_{p,q}$ are independent random complex variables whose real and imaginary pars are independent and have standard Gaussian distribution, with zero mean and variance equal to $1$. The discrete Fourier transform is efficiently performed by the Fast Fourier Transform algorithm. 
    
	\textbf{Percolation clusters.} For a given field realization $\phi(x,y)$ and a level $h$, we consider a non-directed graph whose vertex set is the level set, $V = \{(x,y): \phi(x,y) \le h\}$. Two vertices $(x,y)$ and $(x',y')$ are connected if they are nearest neighbors: $(x',y') = (x\pm 1,y)$ or $(x',y') = (x, y\pm 1)$. The level set clusters are defined the connected components of this graph.
	
	The above procedure defines the site percolation on a square lattice. We can also consider the bond percolation by adding an edge between $(x,y)$ and $(x',y')$, whenever $(x',y') \in \{(x + (-1)^{x+y}, y + (-1)^{x+y} ), (x - (-1)^{x+y}, y - (-1)^{x+y} )\}$ (we assume $L$ is even). The idea is that we view the lattice $\{(x,y)\}$ as that of the middle point of the edges of another square lattice. 
	
	For the figures of this paper, we used the bond percolation. We checked that our results still hold with site percolation.

	\textbf{Method for Figure 1.} We take a \textit{fixed} realization of the complex Gaussians $\{\mathcal{N}_{p,q}\}_{p,q=1}^L$ to generate three fraction GFFs with $H = -1, 0, 1/2$, respectively. Then we plot the largest level set cluster with $h = 0$.

	\textbf{Method for Figure 2.}  For each $L$, we generate $S =10^3$ realizations. For each realization $\phi_i$, we look for the percolation threshold
	\begin{equation}
	    h_i := \sup \{h: \text{The level set} \{(x,y): \phi_i(x,y) \le h\} \text{ has no percolating cluster} \}
	\end{equation}
	by a bisection search (the notion of percolating cluster is defined below). Then the threshold fraction is 
	\begin{equation}
	    f_i := |\{(x,y): \phi_i(x,y) \le h_i\}| / L^2 \,.
	\end{equation}
Then, we plot $P_L(f)$, the empirical cumulative distribution of $\{f_i, i = 1, \dots, S\}$: 
\begin{equation}
     P_L(f) := |\{i=1, \dots, S: f_i \le f\}| / S \,.
\end{equation}

   The percolating cluster is defined as follows. We eliminate the edges that connect a vertex with $y = 1$ to one with $y = L$ (that is, we cut the torus into a cylinder). Then a percolating cluster is one that intersects both boundaries $\{(x,y): y=1\}$ and $\{(x,y): y=L\}$.
	
   \textbf{Method for Figure 4}. For each $L \in \{2^5, 2^6, \dots, 2^{11}\}$, we generated $10^3$ GFF realizations, and determine the level set clusters with $h = 0$. We calculate whether $(x,y)$ and $(x + r, y)$ are in the same level set cluster, for $r = 1, 2, 4, 8, \dots, L / 4$ and all $(x,y)$ in the lattice.  Averaging over $(x,y)$ and realizations gives us a good estimate of $P_2(r)$. 
   
   \subsection{Remarks on the number of encircling loops}
   The analytical arguments in the main text relied on treating the number of encircling loops as deterministic numbers. Let us provide more explanation on this point. 
   
   We first consider the distribution of $d_z$. This can be related to the conformal radii of the interior of the encircling loops viewed from $z$, $\mathrm{CR}(A^z_{k}, z), k = 1, \dots, d_z$, where $A^z_k$ is the interior of the $k$-th loop encircling $z$ (counting from outside); we also define $A^z_0 := \mathbf{D}$ to be the whole domain, so that $ \mathrm{CR}(A^z_{0}, z) \sim L$. In the scaling limit, it is known~\cite{schramm09} that 
   \begin{equation}
     B_k^z := W_{k} - W_{k-1} \,,\, W_k := -\ln \mathrm{CR}(A^z_{k}, z) \,.
   \end{equation}
   are i.i.d random variables with the following generating function 
   \begin{equation}
       \left< e^{\lambda B_k^z } \right> = \sec(\pi \sqrt{2\lambda})
   \end{equation}
   In particular, $B^k_z$ has all the moments, for example,
   \begin{equation}
        \left< B_k^z \right> = \pi^2 \,,\,  \mathrm{Var}(B_k^z) = \frac{5\pi^4}3 \,,
   \end{equation}
   etc. If we view $k$ as time, $W_k$ is a biased random walk with a drift velocity $\pi^2$  which starts at $W_0 \sim -\ln L$. In particular, for large $k$, $W_k$ has a Gaussian distribution centered around $k \pi^2 - \ln L$ and with a standard deviation $\propto \sqrt{k}$. 
   Now, $d_z$ can be determined as the smallest $k$ such that $W_k \ge 0$, i.e., when the loop has a lattice-spacing size $\BigO(1)$. It is then not hard to see that for large $\ln L$, $d_z$ is centered around $\ln L / \pi^2$ (which we found by an independent calculation in the main text) and has a fluctuation $\propto \sqrt{\ln L}$. 
   
    We caution that it is not generally correct to ignore fluctuations of order $\sqrt{A}$ where $A \gg 1$ is the mean value, for example, when calculating the average of $e^{\lambda A}$. Fortunately we do not average over such exponential functions, so our approximation is safe. 
    
    Finally, the above argument can be extended to the quantities $q,s_x, s_y$ involved in the  two-point connectivity. For example, $q$ can be determined as the smallest $k$ such that $W_k \ge -\ln |x-y|$; as long as $\ln L - \ln|x-y| \gg 1$, the argument above carries through, with $\ln L$ replaced by $\ln L - \ln |x-y|$.
	
	\subsection{Fractional Gaussian fields with $H = 0 - \epsilon$}
     A fractional Gaussian field with Hurst exponent $H < 0$ is characterized by the covariance that decays algebraically:
     \begin{equation}
      \left< \phi(x) \phi(y) \right> \sim \frac1{-H} |x-y|^{2H} = \frac1{-H} \exp(2H \ln |x-y|)  \,.
     \end{equation}
     When $H$ is close to $0$, this algebraic decay is indistinguishable from the logarithmic decay of the 2D GFF (which corresponds to $H = 0$) below a crossover scale 
     \begin{equation}
         \ln L_{H} := \frac{1}{-H}  \,.
     \end{equation}
     Indeed, if $1 \ll |x-y| \ll L_H$ we have 
     \begin{equation}
         \frac1{-H} \exp(2H \ln |x-y|) \approx - 2 \ln |x-y| + \frac{1}{-H} = 2 \ln {\frac{L_H}{|x-y|}} \,.
     \end{equation}
     In other words, below the crossover scale, the fraction Gaussian field looks exactly like the 2D GFF in a system of size $L_H$.

	\end{widetext}
	
	\end{document}